\def \SAIT #1 #2 {{\em Mem.\ Soc.\ Astron.\ It.\/} {\bf #1}, #2}
\def \MESS #1 #2 {{\em The Messenger\/} {\bf #1}, #2}
\def \ASTRNACH #1 #2 {{\em Astron. Nach.\/} {\bf #1}, #2}
\def \AAP #1 #2 {{\em Astron. Astrophys.\/} {\bf #1}, #2}
\def \AAL #1 #2 {{\em Astron. Astrophys. Lett.\/} {\bf #1}, L#2}
\def \AAR #1 #2 {{\em Astron. Astrophys. Rev.\/} {\bf #1}, #2}
\def \AAS #1 #2 {{\em Astron. Astrophys. Suppl. Ser.\/} {\bf #1}, #2}
\def \AJ #1 #2 {{\em Astron. J.\/} {\bf #1}, #2}
\def \ANNREV #1 #2 {{\em Ann. Rev. Astron. Astrophys.\/} {\bf #1}, #2}
\def \APJ #1 #2 {{\em Astrophys. J.\/} {\bf #1}, #2}
\def \APJL #1 #2 {{\em Astrophys. J. Lett.\/} {\bf #1}, L#2}
\def \APJS #1 #2 {{\em Astrophys. J. Suppl.\/} {\bf #1}, #2}
\def \APSS #1 #2 {{\em Astrophys. Space Sci.\/} {\bf #1}, #2}
\def \ASR #1 #2 {{\em Adv. Space Res.\/} {\bf #1}, #2}
\def \BAIC #1 #2 {{\em Bull. Astron. Inst. Czechosl.\/} {\bf #1}, #2}
\def \JSQRT #1 #2 {{\em J. Quant. Spectrosc. Radiat. Transfer\/} {\bf #1}, #2}
\def \MN #1 #2 {{\em Mon. Not. R. Astr. Soc.\/} {\bf #1}, #2}
\def \MEM #1 #2 {{\em Mem. R. Astr. Soc.\/} {\bf #1}, #2}
\def \PLR #1 #2 {{\em Phys. Lett. Rev.\/} {\bf #1}, #2}
\def \PASJ #1 #2 {{\em Publ. Astron. Soc. Japan\/} {\bf #1}, #2}
\def \PASP #1 #2 {{\em Publ. Astr. Soc. Pacific\/} {\bf #1}, #2}
\def \NAT #1 #2 {{\em Nature\/} {\bf #1}, #2}

\documentstyle{memsait}
\input epsf.sty

\begin{opening}
\title{The Luminosity Function of 24 Abell Clusters from the CRoNaRio catalogues}
\author{M.Paolillo$^{1,2}$, S.Andreon$^1$, G.Longo$^1$, E.Puddu$^1$, 
S.Piranomonte$^3$, R.Scaramella$^3$, V.Testa$^3$, R.de Carvalho$^4$,
G.Djorgovski$^5$, R.Gal$^5$} 
\institute{$^1$Osservatorio Astronomico di Capodimonte, Napoli, Italy\\
$^2$Dipartimento di Scienze Fisiche ed Astronomiche, Palermo, Italy\\
$^3$Osservatorio Astronomico di Monte Porzio, Roma, Italy\\
$^4$Observatorio Nacional, Rio de Janeiro, Brazil\\
$^5$Department of Astronomy, Caltech, USA}
\date{} 
\end{opening}

\begin{document}

\oddpagefooter{}{}{} 
\evenpagefooter{}{}{} 
\ 
\bigskip

\begin{abstract}
We present the composite luminosity function of galaxies for 24 Abell clusters
studied in our 
survey of the Northern Hemisphere, using DPOSS data in the framework of the
CRoNaRio collaboration. Our determination of the luminosity function has been 
computed with very high accuracy thanks to 1) the use of homogeneous data 
both in the cluster and in the control field, 2) a local estimate of the background,
which takes into account the presence of large-scale structures (which tipically
make the background contribution larger) and of foreground clusters and groups; 
3) the inclusion in the error calculation of the variance of background counts, 
which is tipically larger than Poissonian fluctuation.
We plan at least a tenfold increase of the number of clusters.
\end{abstract}

\section{The CRoNaRio Project (C{\rm altech}-Ro{\rm ma}-Na{\rm poli}-Rio
{\rm de Janeiro})}
The CRoNaRio Project is a joint enterprise among Caltech and the 
astonomical observatories of Napoli, Roma and Rio de Janeiro, aimed to 
produce the first general catalogue of all the objects visible on the DPOSS 
(Digitised Palomar Sky Survey; Djorgovski et al., 1999). The final Palomar-Norris North Sky Catalogue will 
include astrometric, photometric (in the three Gunn-Thuan bands g,r and i) 
and rough morphological information for an estimated number of $2\times 10^9$ 
stars and $5\times 10^7$ galaxies. 

\section{The derivation of Luminosity Functions}
The luminosity function (LF) of galaxy clusters is given by the statistical difference
between galaxy counts in the cluster direction and those in an empty (i.e., not including 
clusters) field.
Therefore, for an accurate computation of the cluster LF we need a good estimate
of the background counts and a careful selection of the surveyed cluster area.

To this end, we take advantage of our wide sky coverage by merging the
CRoNaRio g, r and i catalogues into a ``matched catalogue'' 
(see Puddu et al., this meeting, for more details), containing all those objects which are detected in at 
least two of the three bands, and by deriving a density map of galaxies 
in a $5\times 5$ Mpc (H$_0=50$) region centered on the estimated cluster center.
We then enhance galaxy fluctuations on cluster scales convolving our map with a 
Gaussian function having width corresponding to 250 Kpc in the cluster
rest--frame, {\it id est} to a tipical cluster core radius (Puddu et al.). 
\begin{figure}
\begin{center}
\epsfysize=6.5cm 
\hspace{0.cm}\epsfbox{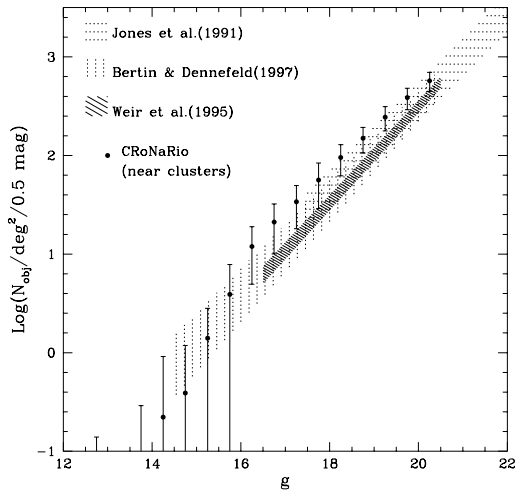} 
\end{center}
\caption[h]{Comparison of near-cluster background counts with some 
previous determinations.}
\end{figure}

In order to estimate the background contribution we first excluded a 3 Mpc wide
circular region centered on the cluster to avoid any unwanted contamination 
from cluster galaxies, and then derived the background galaxy counts and
their {\it variance} over an angular scale comparable to that of the
cluster. 

Such a local correction is more accurate than the traditional average one
for the following reasons:
\begin{enumerate}
\item a local background estimate allows to correct for the underlying 
large-scale structures. These structures are neglected if the average number counts
for the field are used. As shown in Fig. 1, the background near (but not too
near) to the cluster is usually much higher than the average.
\item since local fluctuations are larger than Poissonian errors, our measured
variance correctly estimates the uncertainty introduced by a 
statistical background subtraction.
\item the use of homogeneous data, reduced in the same way, for both the background
and the cluster galaxy counts, allows us to compensate for systematic errors 
due to selection effects, which cancel out (at least in large part) in the
statistical subtraction of the counts.
\end{enumerate}

To detect clusters independently from the center position listed in the Abell
catalog, we searched for the central 1.5 $\sigma$ density peak in 
the inner 3 Mpc circle and then we derived the cluster LF by subtracting, 
from the galaxy counts measured in this region, the background counts 
previously measured, rescaled to the cluster area.
This approach makes it possible to apply the statistical background correction
to the region having higher signal to noise ratio.
Moreover, we estimated the completeness limit of our data for each cluster 
independently, so to take into account the depth variations of our
catalogs from plate to plate and as a function of the cluster location
in the plate.
\begin{figure}
\epsfysize=10cm 
\begin{center}
\hspace{3.5cm}\epsfbox{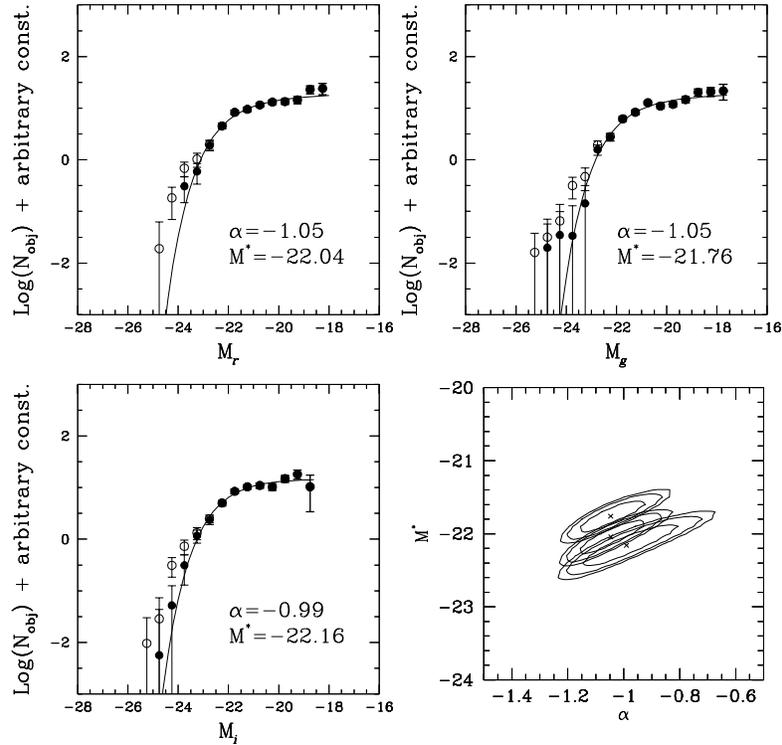} 
\caption[h]{The cumulative LF. Empty dots: brightest cluser member 
(BCM) incuded; filled dots: BCM excluded.}
\end{center}
\end{figure}
\section{The cluster sample}
As a first application of the method, 
we computed the luminosity function of 24 near ($z < 0.28$) and rich Abell
clusters. Redshift were taken from the literature, paying attention to include
in the sample only clusters with well measured redshift.
The composite cluster LF were then cumulated following the Garilli et al.
(1999) method. Figure 2 gives 
the composite LF and the best fitting Schechter (1976) function. 
\begin{figure}
\epsfysize=6.cm 
\begin{center}
\hspace{0.cm}\epsfbox{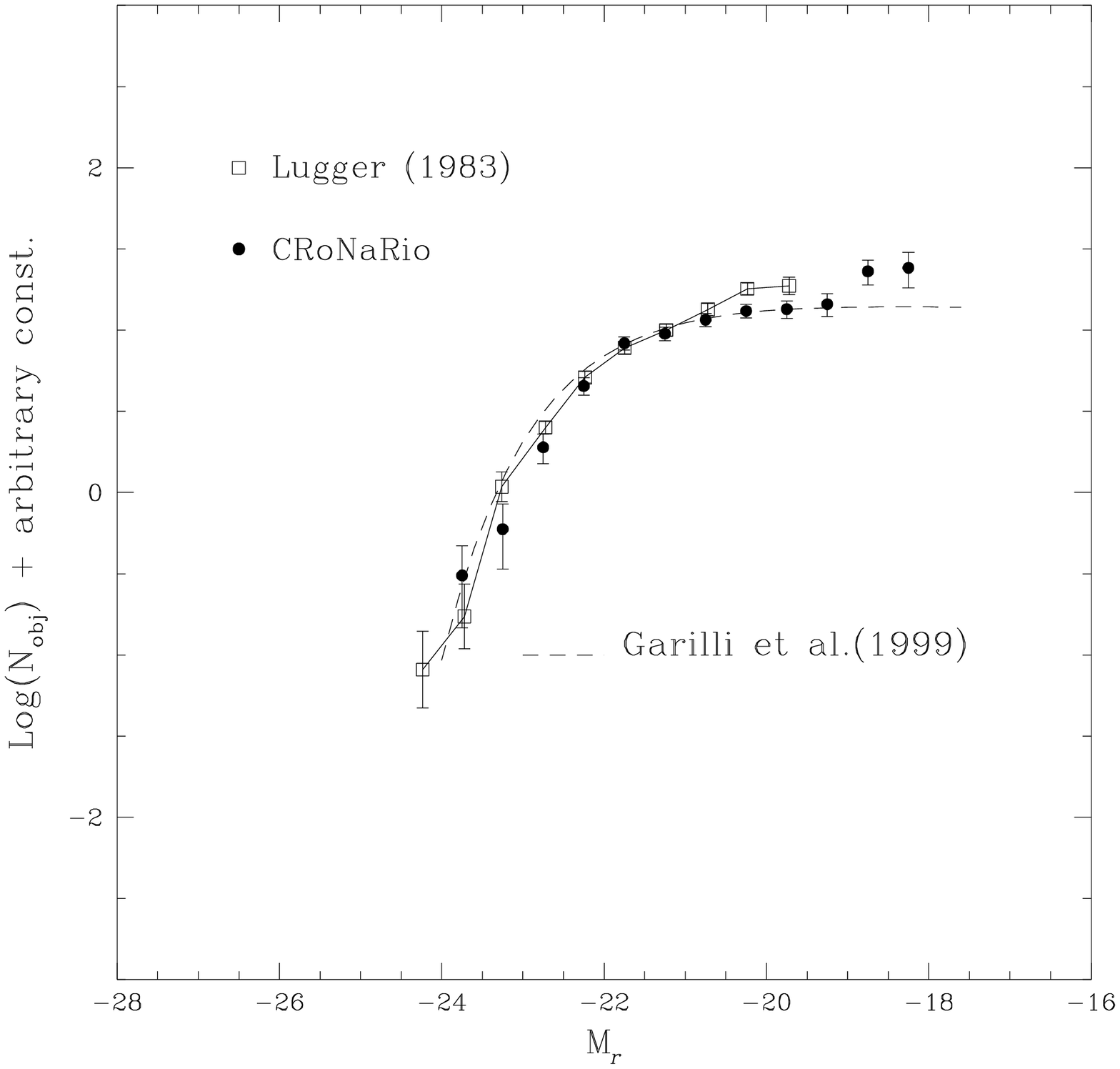} 
\epsfysize=6.cm 
\hspace{0.cm}\epsfbox{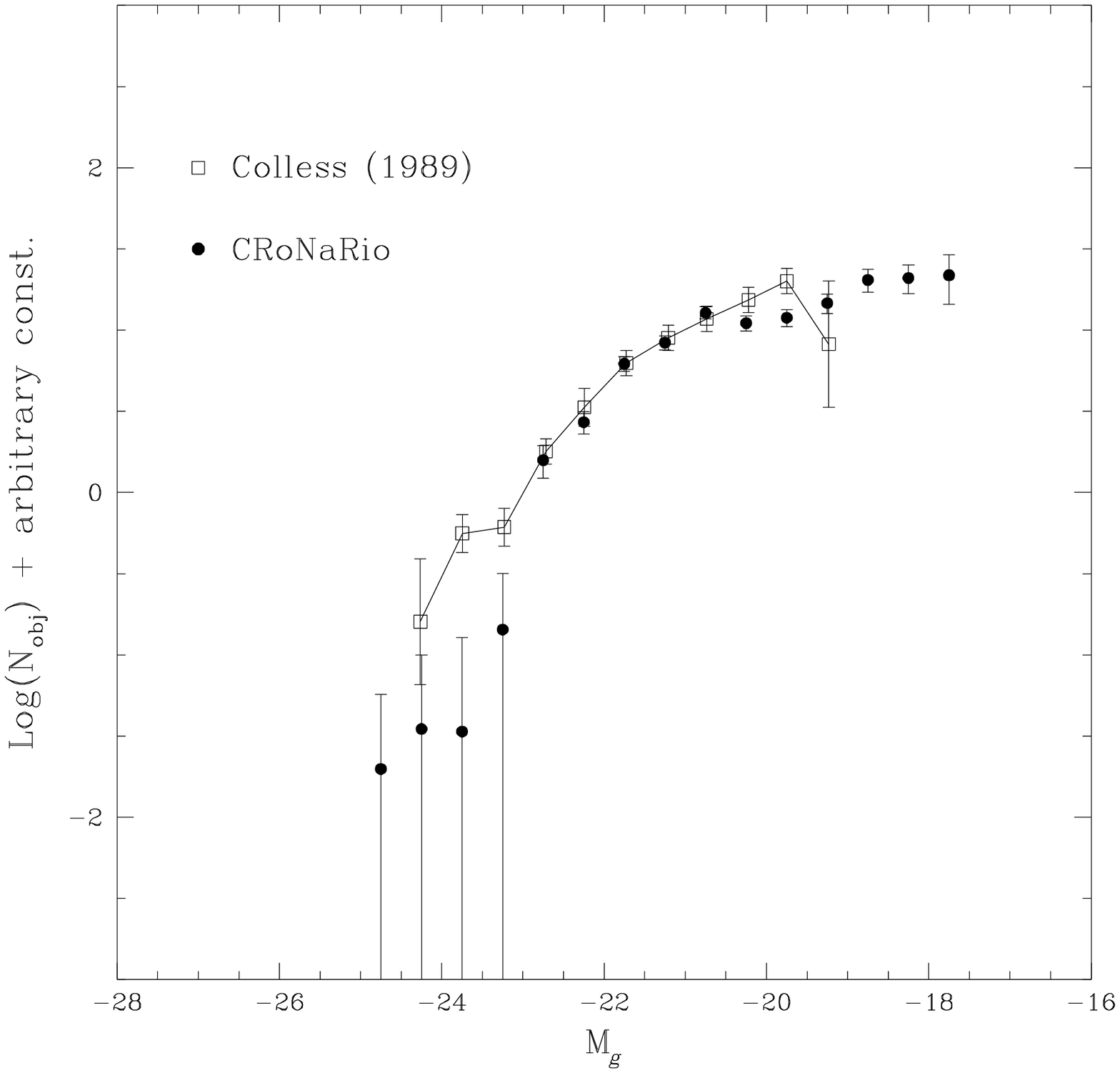} 
\caption[h]{Comparison between the cumulative LF and the LFs obtained by Lugger 
(1983) and Colless (1989) from photographic material, and Garilli et
 al. (1999) from CCD data.}
\end{center}
\end{figure}
\newline\newline
{\bf 4. Discussion of the cumulative luminosity function}\newline\newline
The LF is quite flat: its slope, at M$^*+5$, is $\sim1.0$ in the three filter.
This slope matches well those found by Garilli et al. (1999), based on CCD data, 
who used a completely different method for the background subtraction.
The slope of the LF is flatter than what is traditionally found ($\alpha\sim -1.25$) from
photographic plates by adopting an average background (e.g. Schechter 1976).
Our shallower slope is due to the use of a local background, which is higher
than the average background (see figure 1). 

The LF color, obtained from the differences between $M^*$ values computed in the
three photometric bands, is similar to the colors of early-type galaxies (Fukugita et al., 1995; 
Garilli et al., 1999).

Figure 3 shows how our LF, based on the first 24 clusters of our survey,
matches those available in the literature.
It is important to notice that CRoNaRio data 
reach 1.5 magnitudes deeper than other works based on photographic material
and are comparable to those obtained with CCDs.
 
The huge coverage of the DPOSS - the whole Northern Hemisphere - will allow us
to produce the LF of at least a tenfold higher number of clusters with known 
redshift and to explore the dependence of the LF on the dinamical evolution, 
cluster richness and other environmental and morphological parameters.


\end{document}